\documentclass{aa}
\usepackage{graphicx}

\newcommand{\HI}{{\ion{H}{i}}}

\newcommand{\NII}{{\ion{N}{ii}}}
\newcommand{\SII}{{\ion{S}{ii}}}
\newcommand{\FeII}{{\ion{Fe}{ii}}}
\newcommand{\OIII}{{\ion{O}{iii}}}
\newcommand{\matHI}{\rm H{\hskip 0.02cm\scriptscriptstyle I}}
\newcommand{\abHI}{\rm H{\hskip 0.05cm\scriptsize I}}

\newcommand{\kms}{$\,$km$\,$s$^{-1}$}

\newcommand{\kmsMp}{km s$^{-1}$ Mpc$^{-1}$}
\newcommand{\mJybeam}{mJy beam$^{-1}$}

\def\emph#1{{\sl #1}}
\newcommand{\ltsima} {$\; \buildrel < \over \sim \;$}
\newcommand{\gtsima} {$\; \buildrel > \over \sim \;$}
\newcommand{\lta} {\lower.5ex\hbox{\ltsima}}
\newcommand{\gta} {\lower.5ex\hbox{\gtsima}}

\input{psfig}

\begin{document}

\title{The location of the broad H{\Large\hskip0.1cm I} absorption in 3C305:
clear evidence for a jet-accelerated neutral outflow\thanks{Based on observations
with the Very Large Array.}}

\titlerunning{Broad \abHI\ absorption in 3C~305}
\authorrunning{Morganti et al. }

\author{R. Morganti\inst{1}, T. A. Oosterloo\inst{1},
C. N. Tadhunter\inst{2}, G. van Moorsel\inst{3},
B. Emonts\inst{4}}

\offprints{morganti@astron.nl}

\institute{Netherlands Foundation for Research in Astronomy, Postbus 2,
NL-7990 AA, Dwingeloo, The Netherlands
\and
Dep. Physics and Astronomy,
University of Sheffield, Sheffield, S7 3RH, UK
\and
National Radio Astronomy Observatory, Socorro,
             NM 87801, USA
\and
Kapteyn Astronomical Institute, University of Groningen, P.O. Box 800,
9700 AV Groningen, the Netherlands
}

\date{Received ...; accepted ...}

\abstract{We present high-spatial resolution 21-cm \abHI\ VLA
  observations of the radio galaxy 3C~305 ($z=0.041$). These new
  high-resolution data show that the $\sim 1000$ \kms\ broad \HI\ absorption,
  earlier detected in low-resolution WSRT observations, is occurring against
  the bright, eastern radio lobe, about 1.6 kpc from the nucleus. We use new
  optical spectra taken with the WHT to make a detailed comparison of the
  kinematics of the neutral hydrogen with that of the ionised gas.  The
  striking similarity between the complex kinematics of the two gas phases
  suggests that both the ionised gas and the neutral gas are part of the same
  outflow.  Earlier studies of the ionised gas had already found evidence for
  a strong interaction between the radio jet and the interstellar medium at
  the location of the eastern radio lobe. Our results show that the fast
  outflow produced by this interaction also contains a component of neutral
  atomic hydrogen. The most likely interpretation is that the radio jet
  ionises the ISM and accelerates it to the high outflow velocities
  observed. Our observations demonstrate that, following this strong jet-cloud
  interaction, not all gas clouds are destroyed and that part of the gas can
  cool and become neutral.  The mass outflow rate measured in 3C~305 is
  comparable, although at the lower end of the distribution, to that found in
  Ultra-Luminous IR galaxies.  This suggests that AGN-driven outflows, and in
  particular jet-driven outflows, can have a similar impact on the evolution
  of a galaxy as starburst-driven superwinds.


}
\maketitle

\section{Introduction}

The immediate surroundings of active galactic nuclei (AGN) are complex
regions characterised by extreme physical conditions. There, the
interplay between the enormous amount of energy released from the
nucleus and the ISM is most critical.  

Gas outflows can be a result of such interaction.  Fast nuclear
outflows of {\sl ionised} gas appear to be a relatively common
phenomena in active galactic nuclei (see e.g.\   Crenshaw, Kraemer \&
George\ 2003, Kriss et al.\   2004, Capetti et al.\   1999, Krongold et
al.\ 2003, Veilleux et al.\   2002, Tadhunter et al.\   2001 Elvis 2000).
They are mainly detected in optical, UV and X-ray observations.  Gas
outflows associated with AGN provide energy feedback into the
interstellar medium (ISM) that can profoundly affect the evolution of
the central engine as well as that of the host galaxy (e.g.\  Silk \& Rees
1998, Rawlings \& Jarvis 2004).  The mass-loss rate from these
outflows can be a substantial fraction of the accretion rate needed to
power the AGN.  Thus, they are an important element in the evolution
of the host galaxy.  

It is not too surprising that such outflows are also found in
radio galaxies (see e.g.\ Tadhunter 1991, Tadhunter et al.\ 2001, Holt
et al.\ 2003, van Bemmel et al.\ 2003, and Morganti et al.\ 2004 for a
summary of recent results).  However, it is intriguing that in several
radio sources fast outflows of {\sl neutral} hydrogen (up to 2000
\kms) have been discovered.  The best examples so far are the radio
galaxy 3C~293 (Morganti et al.\ 2003) and the radio-loud Seyfert
galaxy IC~5063 (Oosterloo et al.\ 2000).  The number of galaxies
known to show broad \HI\ absorption (ranging from 800 up to 2000
\kms) is, however, growing and therefore this appears to be a
phenomenon that is relatively common and important in at least some
radio sources (see Morganti, Oosterloo \& Tadhunter 2005 for a
summary).

A number of mechanisms have been suggested to explain these outflows
of neutral hydrogen. They range from starbust-driven superwinds
(Heckman, Armus \& Miley 1990), adiabatically expanded broad emission
line clouds (Elvis et al.\ 2002), dusty narrow-line regions that are
radiation pressure dominated (Dopita et al.\ 2002), to outflows driven
by the radio jet.  Different characteristics, in particular the
location where the outflow is occurring, can be expected depending on
the origin.  For example, if connected to the broad line regions, we
expect to find the outflow located at (or very close to) the
nucleus. If the outflows are driven by the radio jet, we expect an
association between strong radio features and the location of the
outflow. Such an association does not necessarily occur in the
case of radiation driven outflows.   So far, direct information
about the location of the outflow is available only for one object
(IC~5063) and through indirect arguments for a second case,
3C~293 (Emonts et al.\ 2005).  The reason that this information
is available only for such a small number of cases is that high
resolution \HI\ observations performed with a broad enough
observing band as well as  high sensitivity are difficult to obtain.

Here we present results from VLA observations designed to locate the broad
\HI\ absorption in the radio galaxy 3C~305. 3C~305 is a relatively
compact radio galaxy (Heckman et al.\ 1992, Jackson et al.\ 2003) of only
about 5 arcsec (about 4 kpc) in size.  On this small scale, 3C~305 has a
complex structure with two jets forming radio lobes separated by 3.6 arcsec in
PA 54$^\circ$ as seen in MERLIN observations (Jackson et al.\ 2003) as well as
two low-brightness arms extending perpendicular to the radio axis that have
been detected with the VLA (Heckman et al.\ 1982). Deep and relatively narrow
\HI\ absorption was detected in the high-resolution observations done with
MERLIN (Jackson et al.\ 2003). However, recent broad-band Westerbork Synthesis
Radio Telescope (WSRT) observations have revealed that a broad \HI\ absorption
component is also present (Morganti et al.\ in prep.). The spectrum from these
data is shown in Fig.\ 1. The MERLIN observations have a limited bandwidth and
insufficient sensitivity to detect the broad \HI\ absorption component while
at the typical 21-cm WSRT resolution of $\sim 13$ arcsec, 3C~305 appears
spatially unresolved.  Thus, deeper high-spatial resolution {\sl and}
broader-band observations are needed to locate the region where the broad
\HI\ absorption is occurring.

Throughout this paper we will assume $H_\circ= 75$ \kmsMp. At the
redshift of 3C~305 ($z=0.041$) this implies a distance of 167 Mpc, hence 1
arcsec is equivalent to 810 pc.

\begin{figure}
\centerline{\psfig{figure=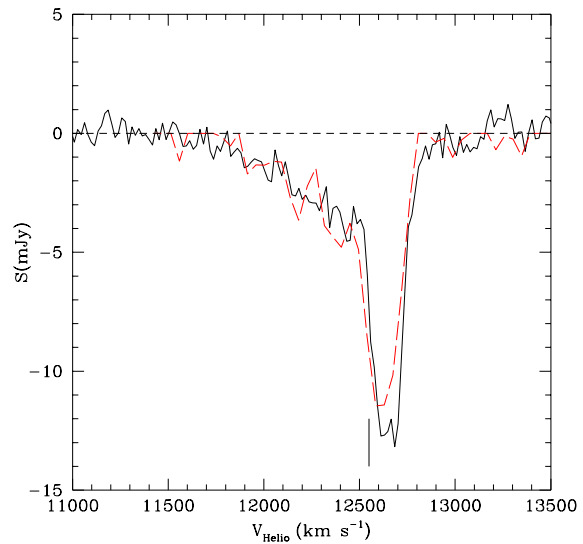,angle=0,width=7cm}}
\caption{\HI\ absorption  profile obtained with the WSRT with a
velocity resolution of $\sim 17$  \kms\ (solid line). The
profile shows a deep, relatively narrow, absorption and a broad
component that covers more  than 1000 \kms\ at zero
intensity. The systemic velocity is also indicated. The long-dashed
profile shows the integrated spectrum from the VLA data (at lower
velocity resolution, see text for details). The similarity of the two
profiles indicates that  basically all the absorbed flux is
recovered from the high-resolution VLA observations.}
\end{figure}
 
\begin{figure}
\centerline{\psfig{figure=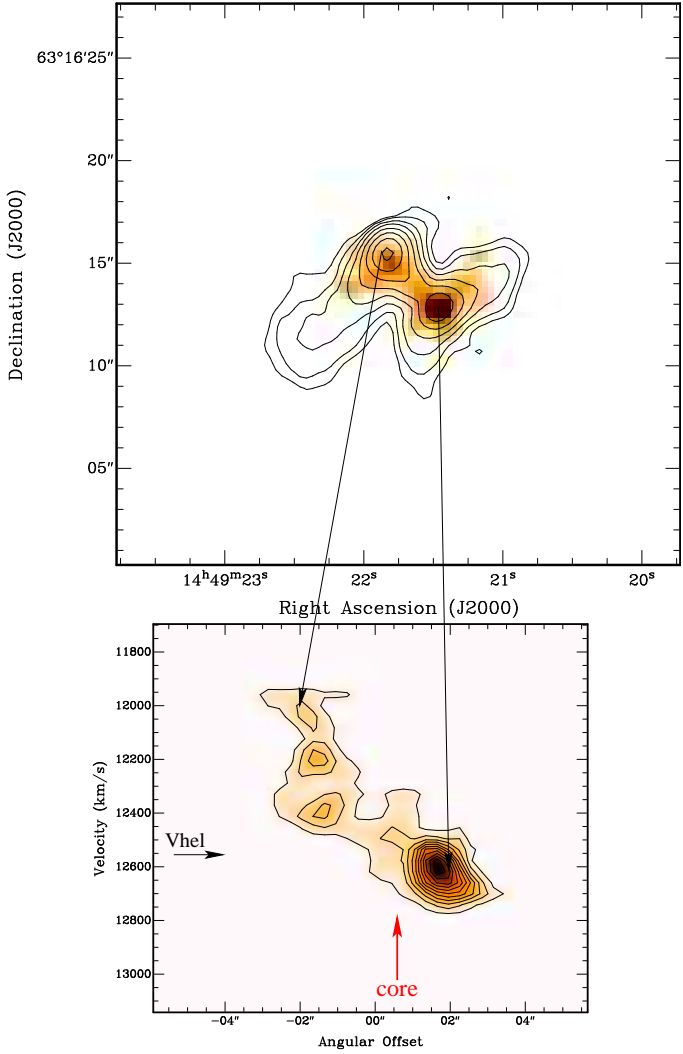,angle=0,width=9cm}}
\caption{Panel showing {\sl (top)} the radio continuum image
 (contour) and the integrated \HI\ absorption  (grey scale). 
  {\sl (Bottom)}
  The position-velocity plot from a slice passing through the two
  lobes.  The contour levels for the continuum image are 10
  \mJybeam\ to 830 \mJybeam\ in  steps of a factor 2.  The
  grey scale image represents the total intensity of the \HI\
  absorption. The contour levels of the \HI\ are $-0.7$, ..., -7.7
  \mJybeam\ in  steps of 0.7 \mJybeam.  The arrow represents the
  systemic velocity.}
\end{figure}

\section{High resolution observations of the \HI\ absorption}

\subsection{VLA observations}

The \HI\ observations were obtained using the VLA in  the A-array
configuration on 23 Sep 2004 with a total integration time on source
of 3.8 h.  The central frequency used was 1363.8 MHz. This frequency
is offset compared to the frequency corresponding to the systemic
velocity of the  galaxy, but corresponds to {\sl the  central
velocity of the entire \HI\ absorption} detected by the WSRT. This
was done in order to make sure that enough continuum  would be
available at both sides of the profile.  The observations made use of
the 12~MHz bandwidth and 64 channels. In order to use this
configuration, we  observed only one polarisation (1 IF). The
velocity resolution obtained is relatively low, only $\sim 40$ \kms\
(before Hanning smoothing), nevertheless good enough for the detection
of the broad component. The need for (at least!) 12 MHz is clear from
the width of the  \HI\ profile.  The data reduction, including
bandpass calibration and continuum subtraction, was done using both
the AIPS and Miriad packages. A line cube was made using uniform
and robust  weighting (using robustness equal to zero). The
results presented  below, and the figures have been obtained for the
robustness equal zero data.  The  beam size is  $1.2
\times 1.0$ arcsec in p.a.$=-18.7^\circ$. The noise per channel (after
Hanning smoothing)  is 0.29 \mJybeam.

From the line-free channels, a continuum image  was made. This
image is shown in Fig.\ 2 (r.m.s.\ noise 1.9 \mJybeam).

\subsection{Results}

In Fig.\ 2 the total intensity of the \HI\ absorption is shown as a
 grey scale image with superimposed contours representing the
continuum image. The \HI\ absorption is  spatially extended and
is detected across the brighter part of the radio source.  The most
important result from these observations can be seen in the
position-velocity plot of Fig.~2, obtained from a slice passing
through the two lobes and the core. The  deep and relatively
narrow part of the \HI\ absorption appears coincident with the SW
radio lobe, as seen before in the MERLIN observations of Jackson et
al.\ (2003).  However, the broad \HI\  absorption appears to be
located in the region of the bright eastern radio lobe,
about 1.6 kpc from the nucleus. This broad component was not seen in
the observations of Jackson et al.\  who used a narrower band.
The broad absorption is weak. Nevertheless, the comparison between the
integrated \HI\ profile from the VLA data with the WSRT profile shows
a great similarity (Fig.\ 1) indicating that  basically all the
absorbed flux is recovered from the high-resolution observations.

The narrow, deep component that covers the SW lobe is spatially resolved and
shows a velocity gradient across the lobe. The optical depth of this component
is $\tau = 0.02$ corresponding to a column density of $N_{\rm \matHI} = 5.4
\times 10^{20} T_{\rm spin}/(100\ {\rm K})$ cm$^{-2}$. This component (centered
on $V_{\rm hel} = 12627$ \kms) was interpreted by Jackson et al.\ (2003) as
due to the nuclear dust lane being located in front of the SW radio lobe and
the SW radio jet pointing away from us. The absence of a similar narrow
component, in their data, against the NE lobe was interpreted as this lobe
being in front of the dust lane and the NE jet pointing towards us.  Our new
observations show that the faint
\HI\ absorption is detected in the region going from the nucleus of
3C~305 up to the bright NE radio lobe. Some of this fainter
absorption could be due to part of the dust-lane being in front of the
central and  some of the NE radio emission. The velocity
gradient near the core could reflect the rotation of the dust-lane
material about the centre of 3C 305. However, the overall \HI\
absorption spans over  such a wide range of velocities (more than
1000 \kms) that not all motions can be due to galactic rotation. In
order to investigate this in more  detail, we will below compare
the velocities of the neutral hydrogen with those of the ionised 
gas (see Sec.\ 3).

The optical depth of the shallow \HI\ absorption is only $\tau \sim 0.0023$ in
the region of the peak of the NE radio lobe and increases along the eastern
jet reaching $\tau \sim 0.01$ at the position of the nucleus. The column
density of the broad component is $N_{\rm \matHI} \sim 2 \times 10^{21}\
T_{\rm spin}/(1000)\ {\rm K} $ cm$^{-2}$. For this component we have assumed a
$T_{\rm spin} = 1000$ K.  The presence of a strong continuum source near the
\HI\ gas, as well as the fact that the gas has likely just passed through a
strong shock, can make the radiative excitation of the \HI\ hyperfine state to
dominate over the, usually more important, collisional excitation (see e.g.\
Bahcall \& Ekers 1969). Under these conditions, the spin temperature is,
therefore, likely to be of the order of 1000 K or more. Using the above
column density, the total \HI\ mass of the outflowing gas can be
estimated. The outflowing neutral hydrogen appears to cover the NE lobe, we
therefore use a region of $1 \times 1$ kpc in size. The resulting mass is
about $1.3 \times 10^7$ $M_\odot$. The size of the region is an uncertain
parameter in this calculation. However, one should also keep in mind that our
observations cannot detect \HI\ located behind of the radio continuum,
therefore more neutral hydrogen could be in principle present in the
region. Thus, the estimated value should give a realistic value to the
\HI\ mass involved in the outflow.

\section{The ionised gas}

In order to obtain more complete information about the kinematics of the gas,
we have investigated the characteristics of the ionised gas using available
long-slit spectra of 3C~305 obtained with the ISIS dual-beam spectrograph on
the William Herschel Telescope (WHT) on La Palma. The wavelength range covers
3300 to 7300\AA\ (in the rest frame of 3C 305), the resolution is 3.6\AA\ (or
$\sim 165$ \kms\ in the red part of the spectrum) and the wavelength
calibration has an accuracy of $\sim 1$\AA\ (or 50 \kms\ at H$\alpha$).  A
more detailed description of the observations and data reduction is given in
Tadhunter et al.\ (2005).

The presence of extended emission lines with complex kinematics was already
known from the work of Heckman et al.\ (1982).  The [\OIII] 5007\AA\ region of
the new spectra (after the subtraction of the galaxy's continuum) obtained
along p.a.\ 60$^\circ$ (the galaxy's major axis) and along p.a.\ 42$^\circ$
(the radio axis) are shown in Fig.\ 3. The location of the peaks of the radio
lobes is marked. For comparison, in the bottom figure, the white contours give
the data from the position-velocity slice (same as Fig.\ 2) of the \HI\
obtained along the same (radio) axis.  From both figures it is immediately
clear how complex and kinematically disturbed the ionised gas is in the region
exactly co-spatial with the bright radio emission, in particular on the
eastern side.

The spectrum taken along the major axis of the galaxy (Fig.\ 3 {\sl
(top)}) shows -  most clearly outside the region of the radio
emission - the signature of  the regularly rotating, large-scale
disk of the  galaxy, with amplitude  of about 400 \kms\ (see
also Heckman et al.\ 1982). The velocity at the location of the peak
of the  optical continuum emission, that we associate with the
nucleus, is 12550 \kms, consistent with the previous measurements of
the systemic velocity of this galaxy (Heckman et al.\ 1982).  In
addition to the rotation, on the eastern side the ionised gas has a
broad and asymmetric  (mostly blueshifted) component, while on the
western side broad redshifted emission is detected. Given the
orientation of the radio source (as discussed in Sect.\ 2.2), this
pattern is indicates a radial outflow of the ionised gas.

In Fig.\ 3 ({\sl bottom})  the spectrum obtained along the radio
axis (p.a.\ 42$^\circ$) is shown, together with the \HI\
position-velocity data. This allows a more consistent comparison
between the kinematics of the neutral hydrogen and the ionized gas.
Along this position angle, the quiescent, rotating gas is still
visible although with a smaller amplitude  because the slit is not
along the major axis. Also in this position angle, a broad and
blueshifted component of the ionised gas is observed  NE of the
core, while in the SW the profiles have a, somewhat narrower,
redshifted wing.

The overlay of the \HI\ position-velocity plot makes clear that the neutral
hydrogen seen in absorption is formed by two components. Some of the \HI\
belongs to the rotating dust-lane structure. As mentioned above, this is the
case for the gas seen against the SW radio lobe and part of this structure
could extend at least to the position of the nucleus and slightly beyond.
Near the NE lobe, the \HI\ profile clearly deviates by almost 500 \kms\ (FWZI)
from the kinematics of the quiescent gas in the galaxy disk - in a similar way
as the broad, blueshifted component of [\OIII].

The interesting result from the comparison in Fig.\ 3 is therefore that the
broad, blueshifted \HI\ is found at the location of maximum disturbance of the
ionized gas on the eastern side region. This strongly suggests that {\sl the
two components of the gas are the result of a gaseous outflow produced by the
same mechanism.}

The kinematics of the ionized gas are illustrated in more detail in Fig.\
4 where the centroid and FWHM of double Guassian fits to the H$\alpha$ line
along the radio axis (p.a.\ 42$^\circ$) are shown.  At all locations two
Gaussians are required to provide an adequate fit to each of the lines in the
H$\alpha$+[\NII] blend, but it is clear from the plot that the splitting
between the two Gaussian components is particularly extreme at radial
distances between 1 and 2 arcseconds NE of the nucleus. This is, therefore,
between the nucleus and the peak intensity of the NE radio lobe. In the
region coincident with the radio lobe there is no evidence for line splitting
in H$\alpha$ but two Gaussians are required to fit the lines, and the broader
of the two components (FWHM$\sim$700 -- 800 km s$^{-1}$) is broader than can
be explained by gravitational motions in a quiescent disk. Note that the
velocity range encompassed by the FWHM of the optical emission lines in the
region of line splitting 0.6 and 1.2 arcseconds to the NE of the nucleus
($\Delta V \sim 1100$ km s$^{-1}$) is significantly larger than the range
covered by the \HI\ absorption at the same location ($\Delta V
\sim 500$ km s$^{-1}$), although the direction of the kinematic disturbance
(blueshift relative to quiescent disk) is the same for both components.

The mass of the ionized gas is directly related to the H$\beta$
luminosity and the density of the gas can be estimated using standard formula
(Osterbrock 1989). We have estimated this mass in the region of the eastern
lobe.  The {\sl total} H$\beta$ luminosity in this region is $1.16
\times 10^{40}$ erg s$^{-1}$.  The measured value of the H$\alpha$/H$_\beta$
ratio ($\sim 3.6$) ths close to the Case B recombination value, suggesting
that the H$\beta$ luminosity is relatively unffected by dust extinction in this
region.  From the ratio of the [\SII]$\lambda$6716/[\SII]$\lambda$6731 lines
we derive an upper limit to the density of 500 cm$^{-2}$.  Using these numbers
we derive a lower limit to the mass 
of the ionized gas in the region of the eastern radio lobe
of $2 \times 10^5 M_{\odot}$. 
Therefore, unless the
actual density is two orders of magnitude less than our upper limit --
improbable given that the gas is likely to have been compressed in a fast,
radiative shock -- this shows that the mass of gas in the ionized
outflow is
much less than that in the neutral outflow.

\begin{figure}
\centerline{\psfig{figure=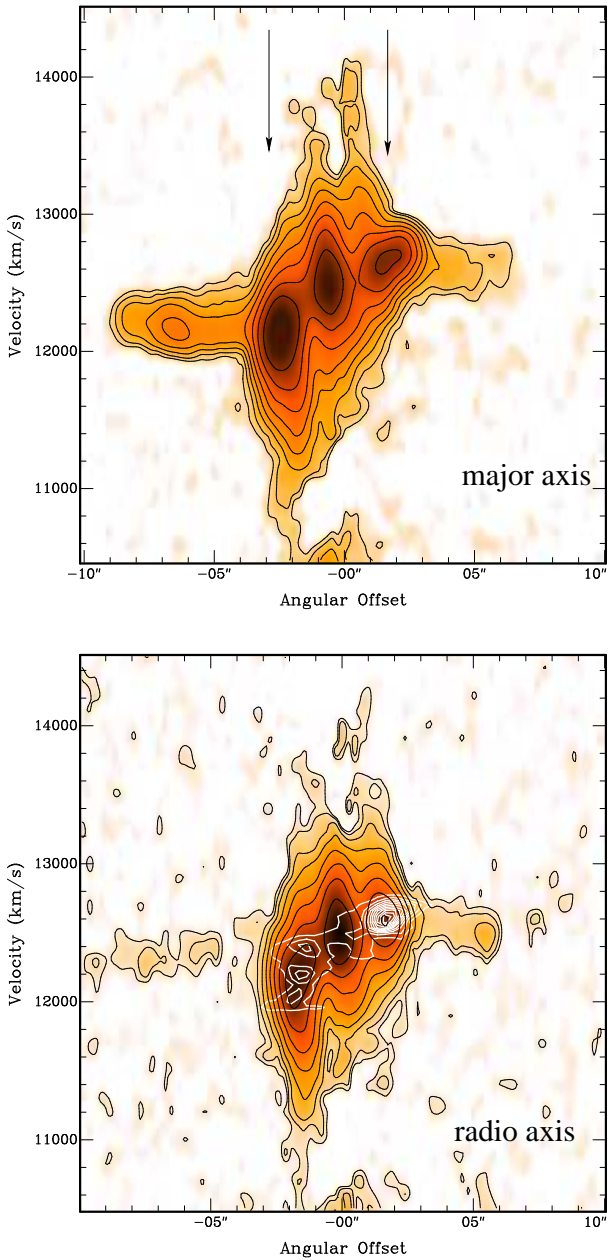,angle=0,width=8cm}}
\caption{{\sl Top} WHT spectrum of the [\OIII] region of 3C~305 (after the
subtraction of the continuum from the galaxy) taken in p.a.\
60$^\circ$, i.e.\ along the galaxy's major axis (NE to 
the left, SW to the right). 
The two arrows
represent the approximate position of the peak of the radio lobes.
{\sl Bottom} WHT spectrum of the [\OIII] region of 3C~305 (black
contours and greyscale) taken in p.a.\ 42$^\circ$, i.e.\ along the
radio axis.  White contours represent the \HI\ position-velocity plot
taken along the radio axis (as in Fig.\ 2).}
\end{figure}

\section{An off-nuclear gas outflow}

The high-resolution {\sl and} the broad-band of new \HI\ observations of the
radio galaxy 3C~305 have allowed us to establish that the blueshifted \HI\
component in this galaxy is located in the region of the NE bright radio
lobe. Compared with the velocities of the quiescent gas in the galaxy disk,
this neutral hydrogen is blueshifted by up to 500 \kms.  This is an important
result as it confirms that neutral hydrogen with very disturbed kinematics is
observed at kpc distances from the nucleus. In the case of 3C~305 the most
blueshifted component of the \HI\ absorption is located at 1.6 kpc from the
nucleus.

While the presence of broad \HI\ absorption is known now for a growing
number of radio sources (Morganti et al.\ 2005), the information about
the location of such absorption is still lacking in most of the cases.
The only exception is the radio loud Seyfert galaxy IC~5063 (Oosterloo
et al.\ 2000) and, through indirect arguments that will need to be
confirmed by high-spatial resolution radio data, in the radio galaxy
3C~293 (Emonts et al.\ 2005).  For these two cases we have argued that
the most likely mechanism to produce the observed outflows of both
ionised and neutral hydrogen is the interaction between the radio jets
and the surrounding (dense) ISM. In the case of 3C~305 the evidence is
also clearly in favour of this explanation.

Evidence of the presence of a strong interaction between the radio plasma and
the surrounding interstellar medium was already obtained in the case of 3C~305
from previous optical studies.  For example, a dense environment has been
suggested to be the cause of the 'H' shaped radio morphology.  It has been
argued, based on the highly disturbed kinematics and outflowing ionised gas
(Heckman et al.\ 1982) as well as the coincidence of [\FeII] emission with the
knot at the end of the NE radio jet (Jackson et al.\ 2003), that the
interaction is particularly strong on the NE side.  The broad, blueshifted
\HI\ is found at the location of maximum disturbance of the ionized gas. This,
therefore, confirms that {\sl fast outflows of neutral hydrogen can be
produced by the interaction between the radio jet and the surrounding dense
medium}.  The presence of neutral gas in this region indicates that the gas
can cool very efficiently following a strong jet-cloud
interaction. Furthermore, it shows that the clouds are not destroyed by this
interaction. This is in agreement with the results obtained by the numerical
simulations used to investigate cases of jet-induced star formation by
Mellema, Kurk \& R\"ottgering (2002) and Fragile et al.\ (2004).

Interestingly, the comparison of the velocities shows that 
the the broad, blueshifted \HI\ absorption does not encompass the the
full range of blueshifted velocities covered by the ionized gas.  To
begin with, one should bear in mind that the blueshifted \HI\ is very
faint and we may be limited by sensitivity in detecting very broad
\HI\ absorption at every location (i.e.\ even if broader components
are present we do not have the sensitivity of detecting them at locations
where the continuum is not as strong as at the peak of the NE
lobe). Nevertheless, there may be some physical reasons for the difference.
The numerical simulations of clouds in radio galaxy cocoons overtaken by a
strong shock wave (Mellema et al. 2002) show that in that scenario the cooling
times for the dense fragments can be very short (only a few times $10^2$
years).  However, the gas accelarated to the highest velocities is of low
density and may not have had time to cool. Alternatively, the higher-velocity
clouds may be destroyed before they cool.

 Another consideration is the geometry of the jet-cloud interaction. This may,
 for example, explain why the highest velocities of the ionized gas are
 observed in the region between the nucleus and the NE radio hot-spot. If the
 acceleration of the gas is due to a bowshock at the location of the hotspot,
 most of the acceleration of the gas at that location may happen in the
 direction perpendicular to the line of sight. On the other hand, further back
 along the jet the observed velocities of the gas are due to the lateral
 expansion of the lobe (mainly along the line of sight). Moreover, only the
 neutral hydrogen located in front of the radio source can be detected, while
 this is not the case for the ionized gas. Hence, the full kinematics of the
\HI\ may not be observable.

The mass outflow rate of the neutral hydrogen is significant. We can
adopt the simple model used in Heckman (2002) and Rupke et al.\ (2002)
to estimate a mass outflow rate.  Following Heckman (2002)

\begin{equation}
\dot{M} = 30\cdot  {{\Omega}\over{4\pi}}\cdot {{r_*}\over{\rm 1\, kpc}}\cdot
{N_{\rm H}\over{10^{21}\, {\rm cm}^{-2}}}\cdot { v \over 300\, {\rm km
s}^{-1}} \ M_\odot\, {\rm yr}^{-1}
\end{equation}

where the mass is flowing into a solid angle $\Omega$ at a velocity $v$ from a
radius $r_*$. Using the column density derived above (N$_{\rm \HI} \sim 2
\times 10^{21}$ cm$^{-2}$) and a mean velocity of the outflow between 200 and
300 \kms, we obtain a mass outflow rate of between 20 and 30 $M_\odot$
yr$^{-1}$ for the neutral gas if we assume that the outflow covers $2\pi$
steradians on the sky, and between 5 and 7.5 $M_\odot$ yr$^{-1}$ if we assume
that the outflow covers only $\pi/2$ steradians.

Thus, the derived outflow rates are higher than those estimated for
the ionised gas (from UV and X-ray observations) in nearby AGN
(Crenshaw, Kraemer \& George 2003). The outflow rate measured in 3C~305
is, instead, comparable, although at the lower end of the
distribution, to that found in Ultra Luminous IR galaxies by Rupke et
al.\ (2002). The outflows observed in those galaxies are related to
starburst-induced superwinds (Heckman et al.\ 1990) and they have been
considered to have a major impact on the evolution of galaxies because
of the feedback effects that these outflows can have (see e.g.\
Veilleux et al.\ 2002, Heckman 2002).  Thus, our result shows that
AGN-driven outflows and in particular jet-driven outflows, can have a
similar amplitude and, therefore, can also have a similar impact on
the evolution of galaxies. This supports the results from numerical
simulations (Di Matteo, Springel, Hernquist 2005) in which the energy
released by the AGN can quench both star formation and the further
growth of the black hole, thus explaining the relationship between the
black hole mass and other properties of the galaxy.  In the case of
3C~305, a significant burst of star formation took place between 0.4
and 1.5 Gyr ago (Tadhunter et al.\ 2005). Despite this stellar burst,
the medium surrounding the radio source is still very dense and only
the radio jet seems to be able to clear this gas.

\begin{figure}
\centerline{\psfig{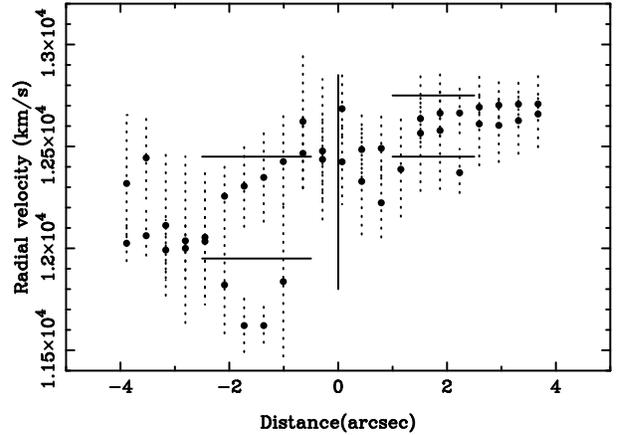}}
\caption{ Plot showing the centroid and FWHM of double Gaussian fits to the
H$\alpha$ line from the spectrum taken along the radio axis (p.a. 42$^\circ$,
NE to the left, SW to the right). The dashed lines represent the range of
velocities encompassed by the FWHM of each Guassian component (corrected for
instrumental broadening). The horizontal lines represent the spatial and
velocity ranges covered by the \HI\ absorption. The vertical solid line
represents the full range of radial velocity covered by the WSRT \HI\
profile. }
\end{figure}

Finally, the \HI\ outflow in 3C~305 may have interesting connections
with the detection of strong and blueshifted \HI\ absorbers (column
density $10^{18} - 10^{19.5}$ cm$^{-2}$) in the Ly$\alpha$ profiles of
high redshift radio galaxies (see e.g.\ Wilman et al.\ 2004 and
ref. therein).  A possible way to explain this is via a highly
supersonic jet expanding into the dense medium of a young radio galaxy
that  then consequently will be surrounded by an advancing
quasi-spherical bow shock, as investigated via numerical simulation by
Krause (2002). The study of the stellar population in 3C~305 suggests
that this galaxy went through a gas-rich, major merger in its recent
past. This radio galaxy may, therefore, represent an ideal example of
nearby object having characteristics very similar to those of typical
high redshift radio galaxies.
 

\begin{acknowledgements}

The National Radio Astronomy Observatory is a facility of the
National Science Foundation operated under cooperative agreement
by Associated Universities, Inc.

\end{acknowledgements}


\begin{thebibliography}{}


\bibitem[]{} van Bemmel, I. M., Vernet J., Fosbury R.A.E., Lamers
H.J.G.L.M. 2003, MNRAS 345, L13
\bibitem[]{} Capetti et al. 1999, ApJ 516, 187
\bibitem[]{} Crenshaw D.M., Kraemer S.B., George I.M.  2003, ARAA 41, 117
\bibitem[]{} Di Matteo T., Springel V., Hernquist L. 2005, Nature in press
(astro-ph/0502199)
\bibitem[]{} Dopita M.A. et al. 2002, ApJ 572, 753 
\bibitem[]{} Elvis M., 2000 ApJ 545, 63
\bibitem[]{} Elvis M., Marengo M. \& Karovska M., 2002, ApJ 567, L107 
\bibitem[]{} Emonts B.H.C., Morganti R., Tadhunter C.N., Oosterloo T.A., Holt
J., van der Hulst J.M. 2005 MNRAS submitted
\bibitem[]{} Fragile, P.C., Murray, S., Anninos, P., van Breugel, W. 2004, ApJ,
604, 74
\bibitem[]{} Heckman T.M., Miley G.K., Balick B., van Breugel W., Butcher
H.R. 1982 ApJ 262, 529 
\bibitem[]{} Heckman T.M., Armus L., Miley G. 1990, ApJS 74, 833
\bibitem[]{} Heckman T.M. 2002, in `Extragalactic Gas at Low Redshift'', eds. J. Mulchaey and J. Stocke, ASP Conf. Series
Vol. 254, p.292  (astro-ph/0107438)
\bibitem[]{} Holt J., Tadhunter C., Morganti R., 2003 MNRAS 342, 227
\bibitem[]{} Krause M., 2002 A\&A 386, L1 
\bibitem[]{} Kriss G. 2004, in IAU Symposium 222 {\it The Interplay among Black
Holes, Stars and ISM in Galactic Nuclei}, eds Storchi-Bergmann et al. in press
(astro-ph/0403685)
\bibitem[]{}Krongold et al.  2003, ApJ 597, 832
\bibitem[]{}Jackson N., Beswick R., Pedlar A., Cole G.H., Spark W.B. et al. 2003 MNRAS 338, 643
\bibitem[]{} Mellema, G., Kurk, J.D., R\"ottgering, H.J.A. 2002, A\&A, 395, L13
\bibitem[]{}Morganti R., Oosterloo T.A., Emonts B.H.C., van der Hulst J.M.,
Tadhunter C. 2003, ApJLetter 593, L69;
\bibitem[]{}Morganti R., Oosterloo T., Emonts  B.H.C., Tadhunter  C.N., Holt
J., 2004, proceedings  IAU Symposium 217, {\em Recycling Intergalactic and
Interstellar Matter}, eds. P.-A. Duc, J. Braine, and E. Brinks ASP, p.332 (astro-ph/0310629); 
\bibitem[]{} Morganti R., Oosterloo T.A., Tadhunter C.N. 2005, in
"Extra-planar Gas", Ed. R. Braun, ASP Conf. Vol 331 in press (astro-ph/0410222) 
\bibitem[]{}Oosterloo T.A., Morganti R., Tzioumis A., Reynolds J., King E.,
McCulloch P., Tsvetanov Z.  2000, AJ 119, 2085
\bibitem[]{}Osterbrock, D.E.~1989, The Astrophysics of Gaseous Nebulae 
and Active Galactic Nuclei, University Science Books, Mill Valley, CA
\bibitem[]{}Rawlings S. \& Jarvis M.J. 2004 MNRAS in press (astro-ph/0409687)
\bibitem[]{}Rupke  D.S., Veilleux S., Sanders D.B. 2002, ApJ, 570, 588

\bibitem[]{}Tadhunter C., Robinson T.G., Gonz\'alez Delgano R.M., Wills K.,
Morganti R. 2005, MNRAS 356, 480
\bibitem[]{}Tadhunter C., Wills K., Morganti R., Oosterloo
T., Dickson R. 2001 MNRAS 327, 227
\bibitem[]{}Tadhunter C. 1991, MNRAS 251, 46
\bibitem[]{}Veilleux S., Cecil G., Bland-Hawthorn J., Shopbell P.L. 2002,
RMxAC, 13, 222
\bibitem[]{} Silk J. \& Rees M.J. 1998 A\&A 331, L1
\bibitem[]{} Wilman R.J., Jarvis M.J., R\"ottgering H.J.A., Binette L. 2004, 
MNRAS 351, 1109



\end{thebibliography}
\end{document}